\documentclass[12pt]{spieman}  
\usepackage{amsmath,amsfonts,amssymb}
\usepackage{graphicx}
\usepackage{setspace}
\usepackage{tocloft}

\title{Electricity Consumption Forecasting for Smart Grid using the Multi-Factor Back-Propagation Neural Network}

\author[a]{Hao Song}
\author[a,*]{Yu Chen}
\author[a]{Ning Zhou}
\author[b]{Genshe Chen}
\affil[a]{Dept. of Electrical \& Computer Engineering, Binghamton University, Binghamton, NY 13902}
\affil[b]{Intelligent Fusion Technology, Inc., Germantown, MD, USA, 20876}

\cftpagenumbersoff{figure}
\cftpagenumbersoff{table} 
\begin{document} 
\maketitle

\begin{center}
\textbf{ABSTRACT}
\end{center}
With the development of modern information technology (IT), a smart grid has become one of the major components of smart cities. To take full advantage of the smart grid, the capability of intelligent scheduling and planning of electricity delivery is essential. In practice, many factors have an impact on electricity consumption, which necessitates information fusion technologies for a thorough understanding. For this purpose, researchers have investigated methodologies for collecting electricity consumption related information and variant multi-factor power consumption forecasting models. In addition, conducting a comprehensive analysis and obtaining an accurate evaluation of power consumption are the premise and basis for a more robust and efficient power grid design and transformation. Therefore, it is meaningful to explore forecasting models that are able to reflect the power consumption changes and internal relations within fusional information effectively. Making electricity consumption forecasting based on the neural network has been a popular research topic in recent years, and the back-propagation neural network (BPNN) algorithm has been recognized as a mature and effective method. In this paper, BPNN is adopted to forecast the electricity consumption using Pecan Street, a community with a relatively large-scale smart grid, as a case study, and takes multiple factors into account, such as weather condition, weekend and holidays. The influences of each factor have been evaluated for a deeper insight. We hope this work will inspire more discussion and further study to guide the design of future smart grids.








\section{Introduction}
\label{sect:intro}  

Electricity consumption prediction has been considered an effective measure that helps the power grid designers and planners build robust, adaptive, efficient, and economic smart grids \cite{zhao2012review}. It is aimed at modeling electricity consumption under different constraints along with environmental factors and the rules. A pre-estimated and calculated electricity demand can be obtained based on the history data including dates, economic, climate and so on. Considering the dynamic pricing mechanism in today’s market, an accurate power load forecasting algorithm is an effective tool for companies to optimize the scheduling load balance decisions to maximize their profit and minimize the probability of accidents like disturbance, overload, etc. Different prediction periods and precision are required for the large scale and complicated smart power grid systems. 

Electricity consumption forecasting is a work which is easy to iterate, but the amount of effort required to improve the quality for each incremental step is huge. The comprehensive consideration of various influencing factors, and the analysis and utilization of diverse types of data to the electricity consumption forecasting models are the requirements of the modern smart power grid. Mastering the way of electricity consumption forecasting with a decent prediction accuracy is a foundation for regional electric power planning, as well as the region's industrial layout, energy distribution, electric power dispatching and power grid investment as a reliable reference.

For better competition in the power market, participants need to accurately predict how much power they will need in a given cycle. On one hand, the underestimation of power demand will lead to higher operation cost \cite{muralitharan2018neural}, which cannot meet the development needs of local economy. On the other hand, overestimation of power demand results in the waste of power resources and investment costs. Therefore, electricity consumption forecasting is one of the most essential tasks in the power market. Electricity consumption prediction can be classified into super-short, short, medium and long term, based on the prediction cycles \cite{al2005long, liu2010short}. Recently, more and more researchers turn their attention to increase the accuracy of the electricity consumption prediction using multiple factors \cite{ahmad2014review, hor2005analyzing}. Because of the limited access of real-world data set, most of the reported works are focused on a single factor or only historical data. There is not much reported research using multiple factors. 

This paper is focused on the effects of different factors that may influence prediction accuracy. For this purpose, leveraging the back propagation neural network (BPNN) algorithm \cite{hecht1992theory} allows multiple factors to be considered and their impacts studied. The accurate analysis and prediction of electrical consumption will help government agencies and the power industry make appropriate electricity utility policies and power scheduling plans. For individual households and the communities on the smart grid, the prediction will help people arrange their electricity usage with more intelligence. 

This paper considers the historical electricity consumption, the weather, and weekend/holiday information to improve prediction accuracy. The effects of different factors are analyzed for a deeper understanding. More specifically, using the historical data of the Pecan Street smart community \cite{barbour2018community} and the corresponding weather information, a BPNN-based prediction has been conducted. The performances under different time resolutions, hourly, daily, weekly, and monthly, are investigated. In addition, the optimal setting of the BPNN is explored according to the data set features by adjusting the parameters to reach the highest performance. 

The rest of paper is organized as follows. In Section \ref{sect:rel}, the research background and related work are introduced. Section \ref{sect:proposal} reviews the principles and design of the BPNN model for power consumption prediction, The Pecan Street data set and its process are discussed in Section \ref{sect:data} and Section \ref{sect:exp} reports the experimental results. Section \ref{sect:conclusions} concludes this paper.

\section{Background and Related Work}
\label{sect:rel}

\subsection{Pecan Street Project}

The Pecan Street Project is an advanced smart community project located in Austin, Texas, USA. Technologies implemented in the participating homes include energy management systems, distributed solar photovoltaic energy, plug-in electric vehicles, smart meters, distributed energy storage, smart appliances, in-home displays, and programmable communicating thermostats \cite{obinna2017comparison}. The Pecan Street Smart Grid maintains over 1,000 households who shared their home or businesses’ electricity consumption data with the project, through the methods such as green button protocols, smart meters, home energy monitoring system and so on. The households in the Pecan Street Project were just like pioneers, they have a great interest in smart community products and services. They have relatively high education and income level in Texas State \cite{obinna2017comparison}. 

Through the Pecan Street Project, massive data can be obtained. Electricity utility data is available with 1-hour resolution, electric data at 1-minute resolution. The gas meter data and the water meter data are collected by encoded radio transmission (ERT). These data have some notable features for researching and modeling: large quantity, high resolution, sustainable access, high reliability and high integrity. It is an ideal candidate for modeling, forecasting and analyzing.

\subsection{Energy Consumption Prediction}

The relationship between the residential electricity consumption and influence factors is often not linear. Quantity analysis-based electricity consumption prediction may get opposite results or unsatisfactory performance \cite{jin2012advances}. Researchers pay more attention to more intelligent algorithms and models. At first, the most popular method was linear regression \cite{almeshaiei2011methodology}. Nowadays, new algorithms like the grey forecasting model, artificial neural network, support vector machine and their corresponded optimization and deformation algorithms become more and more popular and mature \cite{al2005long}.  

Currently, the research on residential electricity consumption is mainly based on household economic theory \cite{samarasinghe2016neural}. Every household purchases electricity corresponding to the household electrical appliance. If there are sufficient data, the model of residential electricity consumption based on household economic theory contains many significant factors, such as electricity price, household income, personal income, alternative energy price, household electrical appliances price, citizens population density, household size, and household area \cite{hahn2009electric, hong2011electric, papalexopoulos1990regression, papalexopoulos1994implementation, park1991electric}. And there are many other factors which may have significant impacts on electricity preference, such as weather or climate condition, holiday, weekend and so on \cite{hippert2001neural}. 

In some developed countries and areas, such as Europe, the United States, Japan, Hong Kong, there are many studies on multifactor residential electricity consumption models \cite{akay2007grey, ekonomou2010greek, kavaklioglu2009modeling, li2015building}, due to the powerful data collection work system. But most of these researches only consider one or few of the factors mentioned above. Some researches even only focused on historical records of electricity consumption and ignore all other factors. Some researchers have done multiple indicators annual prediction. Most of them introduced the factors of installed power capacity, historical yearly electricity consumption, gross domestic product, popularity, imports, exports and so on \cite{ekonomou2010greek, kavaklioglu2009modeling, kaytez2015forecasting}. Some of them only consider historical consumption data \cite{akay2007grey}. According to the existing records and experimental results, the annual prediction is suitable for an extra-large zone (e.g. a country). And when it is accessible to obtain data like gross domestic product and popularity, the artificial neural network (ANN) is mostly applied, if not, the grey model can take its advantages \cite{akay2007grey, hamzacebi2014forecasting}.

For short-term and medium-term predictions, the ANN is also widely used. Many researchers have done optimizations based on the neural network itself. GA, PSO and Elman Neural Network are some examples that have good performance \cite{li2015building, beccali2008short, azadeh2007integration, azadeh2008simulated, escriva2011new}. After optimization, the training speed, the prediction accuracy become better at certain degree. 
Some researchers have made changes on input factors or indicators, for example, introducing temperature, weekday or weekend, seasons \cite{tso2007predicting, neto2008comparison, gonzalez2005prediction}. According to some experimental results, changes of the number of historical consumption data will affect the prediction performance \cite{karatasou2006modeling, park1991electric, taylor2002neural, mandal2006neural}, like the results of 1 hour before input and 24 hours input are different. Also, for the structure and parameters of the neural network, there are more spaces to adjust, like the number of layers, the number of neurons in each hidden layer, the learning rules, the transfer function between each layer and so on \cite{ekonomou2010greek, kaytez2015forecasting}.

Besides ANN and Grey Model (GM), there are many other algorithms and models, such as Support Vector Machine (SVM), regression analysis, detail model simulation, statistical methods, and decision tree and so on. There are many comparisons among these methods \cite{ekonomou2010greek, tso2007predicting, kaytez2015forecasting, neto2008comparison, zhao2012review, darbellay2000forecasting}. For different cases, different areas, different data set types, these methods have different performance and adaptation, and different methods have different model complexity, usability, running speed, input needs and accuracy. For example, in principle, the regression analysis does not have higher accuracy than ANN, but its model complexity is lower than ANN, sometimes the cost performance ratio is an important consideration basis. 

\subsection{BP Neural Networks}

Neural networks are a complex nonlinear system that consists of numerous neurons. In this system, every neuron has a relatively simple function and construction. However, when they are merged together into the entire system, the behavior can be very complex. In artificial neural networks, strength and condition of every connection between nodes are adjustable, it has strong ability of self-learning and self-adaption. The artificial neural network can be applied to many aspects and research areas. Dividing the data samples into three parts: the training data set, the validation data set and the testing data set. The training data set and the validation data set are used in the process of ``training'', and the validation data set is randomly picked up from training data set in some proportion. 

In an artificial neural network, the neurons can be classified into three types according to their position and the information they process: input units, hidden units and output units. The input units receive the input information of the system, which represent the outside signals or data. The output unit give the output after neural network processes, which represent the result. Hidden units form a layer between the input units and the output units. While they do not represent any information of the entire system, but they are significant to the entire neural network and have profound impact on the prediction results. The connection between each neuron mainly reflect the process of information from input to output. This process is repeated many epochs, it is the important part of artificial neural network learning and training. 

The BPNN is one of the artificial neural networks that are widely used in many research areas. This technique is also sometimes called backward propagation of errors because the error is calculated at the output layer and feedback through the network layers. There are two main processes in BPNN learning: The first one is propagation, which includes the generation of the output from each layer and the error (the difference between actual output and target value); the second one is updating the weight. 

For weight update, multiply the error of weight’s output and input activation, then find the gradient of the weight. A ratio of the gradient of the weight is subtracted from the weight. This ratio is named as learning rate which can affect the training speed and performance. If the learning rate is low, the training will become more reliable, but the optimization will take a long time because each step of the minimum value of the orientation loss function is small. If the learning rate is high, the training may not converge at all or even spread out. The change in weight can be so large that the optimization goes over the minimum, making the loss function worse. The weights need to be updated in the opposite direction of the gradient, thus this method is called gradient descent.

\section{BPNN for Power Consumption Prediction}
\label{sect:proposal}

The BPNN algorithm is one of the most widely used ANN models. It is a multi-layer feedforward network and its key feature is back-propagating the error. It is applied to learn and memorize huge amount of mapping relations of input-output models, and there is no need to disclose in advance the mathematical equation that describes these mapping relations. Its learning rule is to adopt the steepest descent method, where the back propagation is used to regulate the weight value and threshold value of the network to achieve the minimum error sum of square.

\begin{figure} [ht]
\begin{center}
\begin{tabular}{c}
\includegraphics[height=5.5cm]{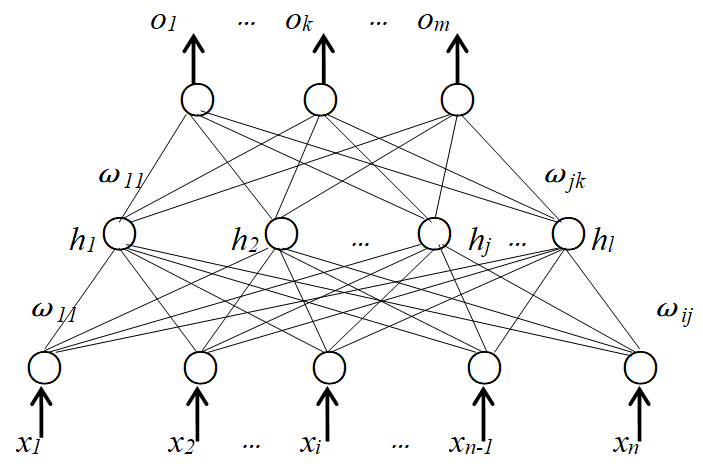}
\end{tabular}
\end{center}
\caption[example] {Structure of BP Neural Network.}
\label{fig:bpnn} 
\end{figure}


Figure \ref{fig:bpnn} shows a basic BPNN architecture, in which $x_1, x_2, \ldots, x_n$ are the neurons in the input layer, $h_1, h_2, \ldots, h_n$ are the neurons in the hidden layer, $o_1, o_2, \ldots, o_n$ are the neurons in the output layer. $ω_{ij}$ is the weight from neuron $i$ in the input layer to neuron $j$ in the hidden layer. $ω_{jk}$ is the weight from neuron $j$ in the hidden layer to neuron $k$ in the output layer. 




Since BPNN has been thoroughly discussed in literature \cite{fausett1994fundamentals, paola1995review}, this section presents the rationale of parameter selection that serves our purpose. In this paper, the initial weight and bias of each layer are randomly picked by the MATLAB neural network tool box. The larger the learning rate, the quicker the learning processes; however, quicker learning processes have lower accuracy. To get a balance between learning speed and accuracy, the learning rate is selected as 0.01. 
There is one node in the output layer and it reflects the predicted electricity consumption. 
The number of input layer nodes will vary depending on various factors. For example, if the historical electricity consumption is the only factor considered, there will be fewer nodes examined than when other factors are taken into account, such as the weather and weekend/holiday. For the number of nodes in the hidden layer, two necessary conditions are considered:

\begin{enumerate}
    \item The number of nodes in the hidden layer must be fewer than $N-1$. Here $N$ is the number of training samples. Otherwise, the error of the network model will have no relationship with the training samples’ features and it will race to zero. Thus, there is not a generalization ability of the network model; and
    \item The number of training samples should be larger than the connection weight of the network model.
\end{enumerate}
	 
According to these two conditions and previous research \cite{karsoliya2012approximating}, there are four empirical formulas to quantify the hidden layers. The first one is:

\begin{equation}
   \label{eq:exp1}
    \sum_{i=0}^{n} C_{m}^{i} > k
\end{equation}

\noindent{where $k$ is the number of training samples, $m$ is the number of nodes in hidden layer, $n$ is the number of nodes in input layer, and $i$ is a constant value locate in $[0, n]$.} The second equation is:

\begin{equation}
   \label{eq:exp2}
     n_1 = \sqrt{n+m}+a
\end{equation}

\noindent{where $n_1$ is the number of nodes in the hidden layer, $n$ is the number of nodes in the input layer, $m$ is the number of nodes in the output layer, and $a$ is a constant value locate in $[1, 10]$. The third and fourth equations are as follows:}

\begin{equation}
   \label{eq:exp3}
     n_1 = \log_2n
\end{equation}

\begin{equation}
   \label{eq:exp4}
     n_1 = 2n+1
\end{equation}

\noindent{where $n_1$ is the number of nodes in the hidden layer and $n$ is the number of nodes in the input layer.}

In the actual implementation of experimental study, the trial and error method will be used to test these empirical formulas in order to find the optimal number of nodes in the hidden layer. Details will be discussed in Section \ref{sect:exp}.

\section{Data Preparation}
\label{sect:data}

\subsection{Data Acquiring}

The historical electricity consumption data is downloaded from the Pecan Street Project, as collected by Pecan Street Inc. (www.pecanstreet.org). It provides the hourly electricity consumption data in its smart grid community. For this paper, ten households of electricity consumption were collected. These ten households were randomly picked up from the community. These households have been offering their power consumption information for many years. As such, it is a stable and reliable source of data, which makes it convenient for future research and validation. Considering the size of the data set, two years (2016 and 2017) data are collected, which show the households’ hourly electricity consumption. The data has more than 17,000 record points which meet the requirement for the BP neural network prediction model. The weather condition data of Austin, Texas was downloaded from the National Climatic Data Center (NCDC). In this paper, two factors were considered: temperature and humidity. Weekend information was gathered through observing general calendar trends. Because different states have different holidays, holiday information was taken from the Office Holidays of Texas State (www.officeholidays.com ). Thus, the raw data set has three main parts: historical electricity consumption, weather information and weekend/holiday information. 

\subsection{Erroneous and Missing Data Processing} 

The raw data from real-world is not perfect, there are two issues need to be addressed, missing data and outliers. There are a few data errors where the information was recorded incorrectly. For example, the temperature or humidity was recorded as -999.99 for a given hour. To deal with this kind of error, the average value of two hours before the recorded error and two hours after the recorded error is used. In addition to data errors, there are many data points missing. While some missing data consists of a single hour, in some cases, there are several continuous hours missing. Theoretically speaking, there should be 17,544 records for hourly electricity consumption for 2016 and 2017. However, the raw data only contains 17,427 records. Even if the missing data consists of less than 1\% of the entire data set, it will bring negative effects on the final prediction accuracy. Since the power consumption information is sequential, any missing data will break the consecutiveness of information. For example, if data from 1 pm to 7 pm is missing, the MATLAB program will automatically read 7 pm data to fill in 2 pm data slot and so on. When an individual hour record is missing, the error is resolved in a manner similar to correcting data errors, using the average value of two hours before the missing value and two hours after the missing value. If the missing data covers several continuous hours, a different method is applied. For example, the power consumption records between 07/08/2016 and 07/09/2016 are incomplete because many hourly data records are missing. In this case, it is not feasible to simply calculate the average value of two hours before and two hours after because so many hours of data are missing. Instead, the data of two days before the missing value range and two days after the missing value range that cover the missing data range are considered. 

\subsection{Data Categories}

In order to achieve a higher prediction accuracy, the data set has been adjusted accordingly depending on different prediction scales: hourly, daily, weekly and monthly. For hourly prediction, the data set has 17520 rows and eight columns. Rows represent the hourly time (for example 6:00, 7:00 and so on). The eight columns follow the given format: electricity consumption (in kWh), month, temperature (in Fahrenheit), humidity (in \%), hour (which hour in a day), day (which day in a week), whether or not it is a weekend, and whether or not it is a holiday.

For the daily prediction, the data set has 730 rows and nine columns. Each row corresponds to a day. The nine columns follow the following format: electricity consumption (in kWh), the highest temperature of the day (in Fahrenheit), the lowest temperature of the day (in Fahrenheit), the average temperature of the day (in Fahrenheit), the highest humidity of the day (in \%), the lowest humidity of the day (in \%), the average humidity of the day (in \%), whether or not it is a weekend, and whether or not it is a holiday.

For the weekly prediction, the data set has 104 rows and eight columns. Each row represents a week. The eight columns follow the following format: electricity consumption (in kWh), the highest temperature of the week (in Fahrenheit), the lowest temperature of the week (in Fahrenheit), the average temperature of the week (in Fahrenheit), the highest humidity of the week (in \%), the lowest humidity of the week (in \%), the average humidity of the week (in \%), and the number of holidays in the week.

For the monthly prediction, the data set has 48 rows and nine columns. Each row represents a month. The nine columns follow the following format: electricity consumption (in kWh), the highest temperature of the month (in Fahrenheit), the lowest temperature of the month (in Fahrenheit), the average temperature of the month (in Fahrenheit), the highest humidity of the month (in \%), the lowest humidity of the month (in \%), the average humidity of the month (in \%), the number of weekends in the month, and the number of holidays in the month.

The daily, weekly and monthly data are actually statistics of the hourly data. For example, to calculate the daily data set, the electricity consumption data is obtained by adding the total hourly electricity consumption on that day. The highest and lowest temperatures are found among the hourly reports for the day. The average temperature is the mean value of the 24-hour period. Calculating daily humidity follows the same process as calculating daily temperature. The weekend and holiday information can be checked using the calendar. 

To calculate the weekly data set, the electricity consumption data is obtained by adding the total hourly electricity consumption for that week. The highest and lowest temperatures are found among the hourly reports for the week. The average temperature is the mean value of the 168-hour period. Calculating weekly humidity follows the same process as calculating weekly temperature. Because there are always two weekend days in a week, thus weekend information is meaningless in the weekly data set. The holiday information can be checked using the calendar.

For the monthly data set, the electricity consumption data is calculated by adding the total hourly electricity consumption for that month. The data cannot be pulled from weekly reports because a given month does not always have the same number of weeks as another month. For example, the first part of a week could belong to June and the rest of the week could belong to July. The highest and lowest temperatures are found among the hourly reports for the month. The average temperature is the mean value of the hourly records for the month. Calculating monthly humidity follows the same process as calculating monthly temperature. The weekend and holiday information can be checked using the calendar.

Following these data processing steps, the possible errors in the obtained four data sets are reduced.

\subsection{Data Normalization}
In the machine learning area, different evaluation indicators have different measurement units and order magnitude, these indicators refer to each column in the data set discussed in the previous subsections. The original data with original measurement units and magnitude will make it hard to get satisfactory analysis result and training performance. To reduce this effect, a standardization process is necessary and important. Among many standardization methods, the normalization process is one of the most typical approaches. 
In this paper, the input metric is normalized by mapping raw minimum and maximum values to [-1,1]. Using processing instruction \textit{mapminmax} in MATLAB tool box, which normalizes the data set row by row. 

\section{Experimental Study}
\label{sect:exp}


\subsection{Experimental Settings}

There are two methods for obtaining hourly prediction: with or without the weather and weekend/holiday factors in the data set. When omitting the weather and weekend/holiday factors, the inputs only contain the historical electricity consumption data. To find how many historical data inputs should be taken into consideration to achieve the highest accuracy and lowest mean squared error (MSE) for hourly prediction, several tests were conducted by considering: 

\begin{itemize}
    \item 0 hour (only considering weather and weekend/holiday information); and
    \item 1, 2, 4, 6, 12, and 24 hours (with \& without weather and weekend/holiday information);
\end{itemize}

For daily prediction, several tests were conducted following a pattern similar to hourly prediction by considering: 

\begin{itemize}
    \item 0 day (only considering weather and weekend/holiday information); and 
    \item 1, 3, 5, 7, 9, 11, and 13 days (with \& without weather and weekend/holiday information).
\end{itemize}

For weekly prediction, several tests were conducted by considering: 

\begin{itemize}
    \item 0 week (only considering weather and weekend/holiday information); and
    \item 1, 2, 3, 4, and 5 weeks (with \& without weather and weekend/holiday information).
\end{itemize}

For monthly prediction, several tests were conducted by considering: 

\begin{itemize}
    \item 0 month (only considering weather and weekend/holiday information); and
    \item 1, 2, 3, and 4 months (with \& without weather and weekend/holiday information).
\end{itemize}

For the different tests above, the number of neurons in the input layer and the hidden layer must be adjusted correspondingly. 

\subsection{Experimental Results}

\begin{figure} [htb]
\begin{center}
\begin{tabular}{c}
\includegraphics[height=4.5cm]{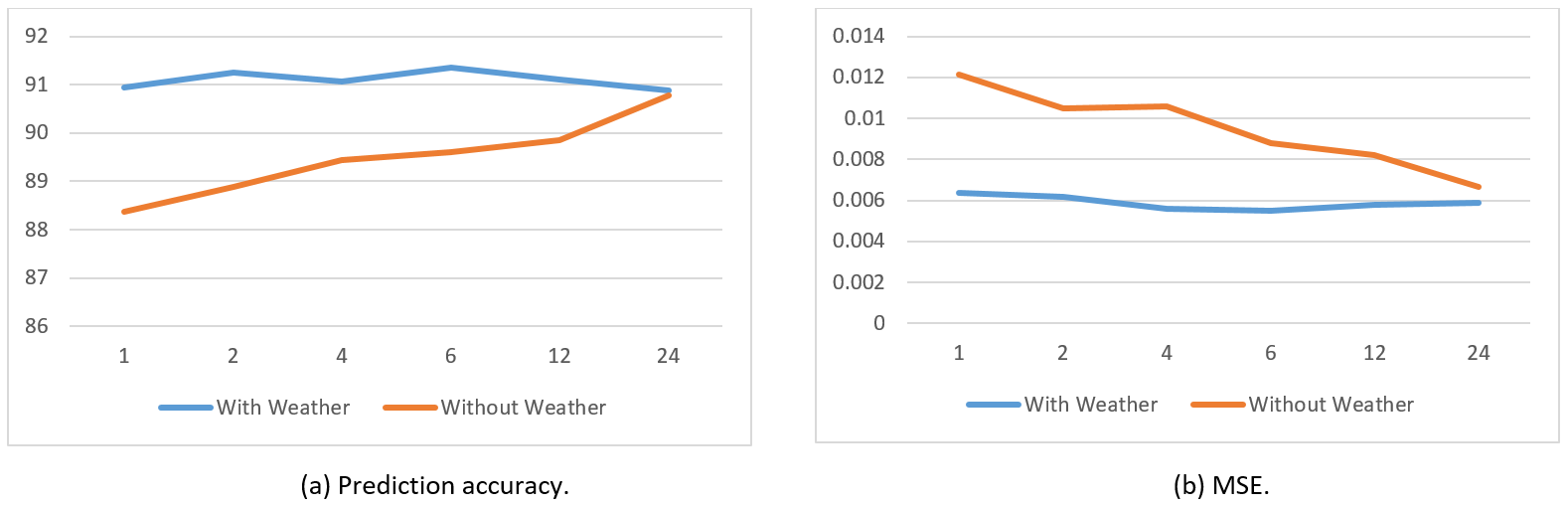}
\end{tabular}
\end{center}
\caption[example] {Accuracy and MSE of Hourly Predictions.}
\label{fig:hourly} 
\end{figure}

Figures \ref{fig:hourly}(a) and (b) demonstrate the influences of historical electricity consumption and the weather and weekend/holiday information on the prediction accuracy. When weather condition and weekend/holiday information are not considered, a longer history record helps to achieve higher accuracy and lower error. In contrast, when the weather and weekend/holiday information is included, the prediction accuracy is higher and it becomes insensitive to the length of the historical record. In general, the history information is useful to improve the prediction accuracy, much fewer historical records is needed to achieve the decent level of accuracy when the weather and weekend/holiday factors are available. This actually implies lower computing and transmission overhead. 

\begin{figure} [htb]
\begin{center}
\begin{tabular}{c}
\includegraphics[height=4.5cm]{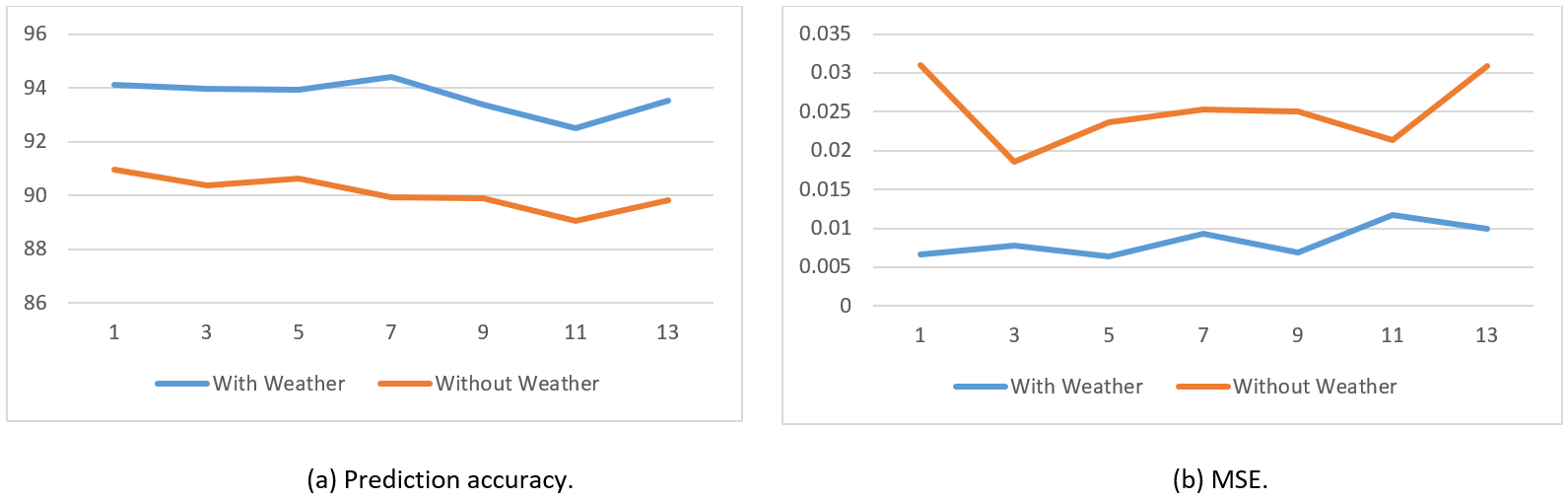}
\end{tabular}
\end{center}
\caption[example] {Accuracy and MSE of Daily Predictions.}
\label{fig:daily} 
\end{figure}

In terms of daily prediction, the results are shown in Fig. \ref{fig:daily}. There is not a clear relationship between prediction performance and the amount of historical electricity consumption data applied. However, it may be an experience that using seven days of historical record has the highest accuracy with a fair MSE. When the weather and weekend/holiday factors are not considered, it looks the accuracy is lower when longer historical record is applied, and the MSE is not related to the length of historical records. As a conclusion, in daily prediction, considering weather and weekend/holiday factors yields much higher accuracy and lower MSE but the longer historical record does not improve the prediction accuracy.

\begin{figure} [htb]
\begin{center}
\begin{tabular}{c}
\includegraphics[height=4.5cm]{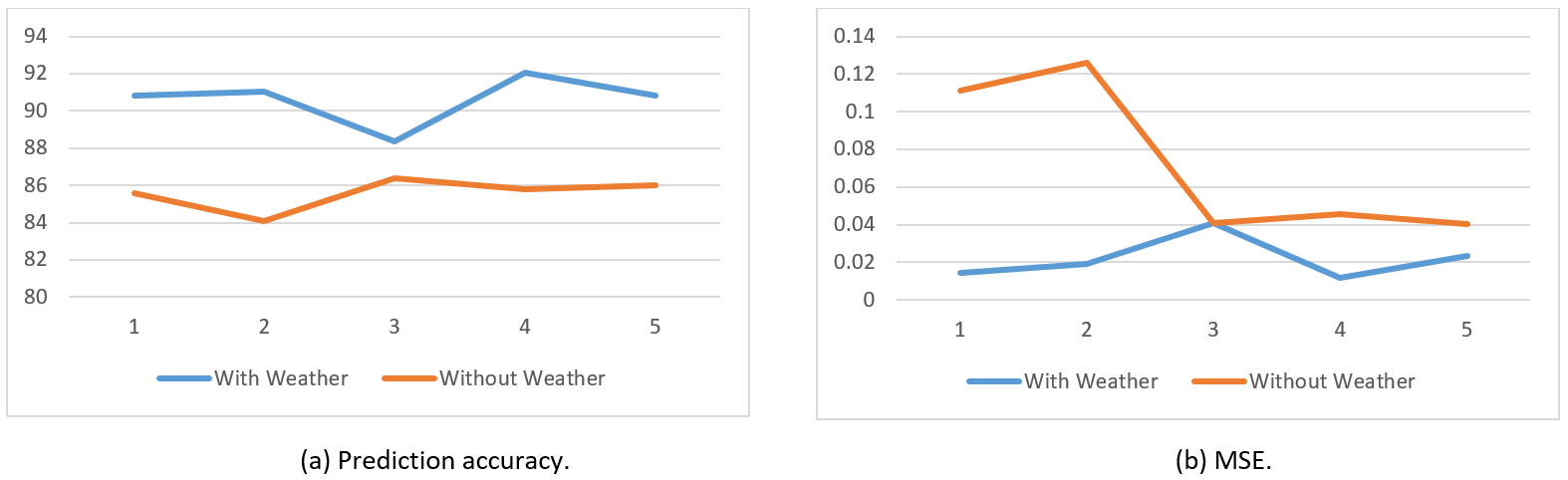}
\end{tabular}
\end{center}
\caption[example] {Accuracy and MSE of Weekly Predictions.}
\label{fig:weekly} 
\end{figure}

Figure \ref{fig:weekly} demonstrates that for the weekly prediction, when the weather and weekend/holiday factors are considered, including more weekly historical electricity consumption data actually decreases the prediction accuracy. When the weather and weekend/holiday factors are not considered, more weekly historical electricity consumption data does not always contribute to the predictions. Using three weeks of historical data got the best prediction with the highest accuracy and the lowest MSE. But two weeks of the historical data leads to the worst result. 

\begin{figure} [htb]
\begin{center}
\begin{tabular}{c}
\includegraphics[height=4.5cm]{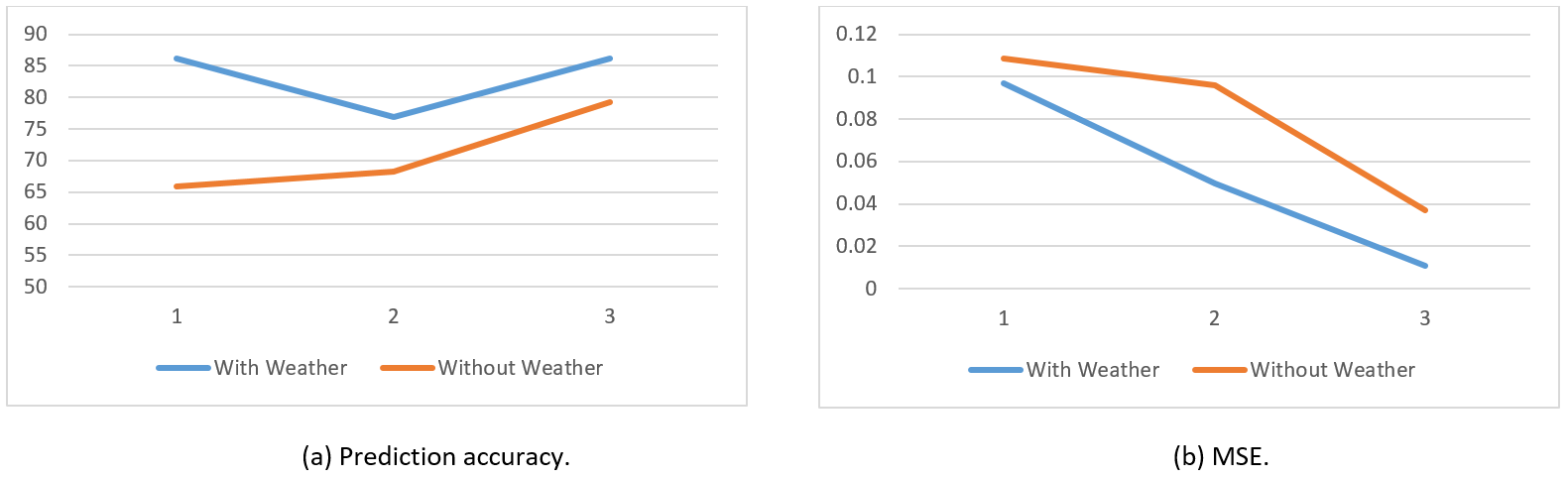}
\end{tabular}
\end{center}
\caption[example] {Accuracy and MSE of Monthly Predictions.}
\label{fig:monly} 
\end{figure}

Figure \ref{fig:monly} demonstrates the change of prediction performance for the monthly prediction with weather information and weekend/holiday factors included. The accuracy decreases when more monthly historical electricity consumption data is used. When two months of historical data is used, the accuracy is the lowest. The performance recovers after that, however, due to the limited data set, this work could not try more. It is very interesting for monthly prediction, when the weather information and weekend/holiday factors are not considered, the prediction performance will improve as more historical electricity consumption data inputs are added.

\begin{figure} [htb]
\begin{center}
\begin{tabular}{c}
\includegraphics[height=12cm]{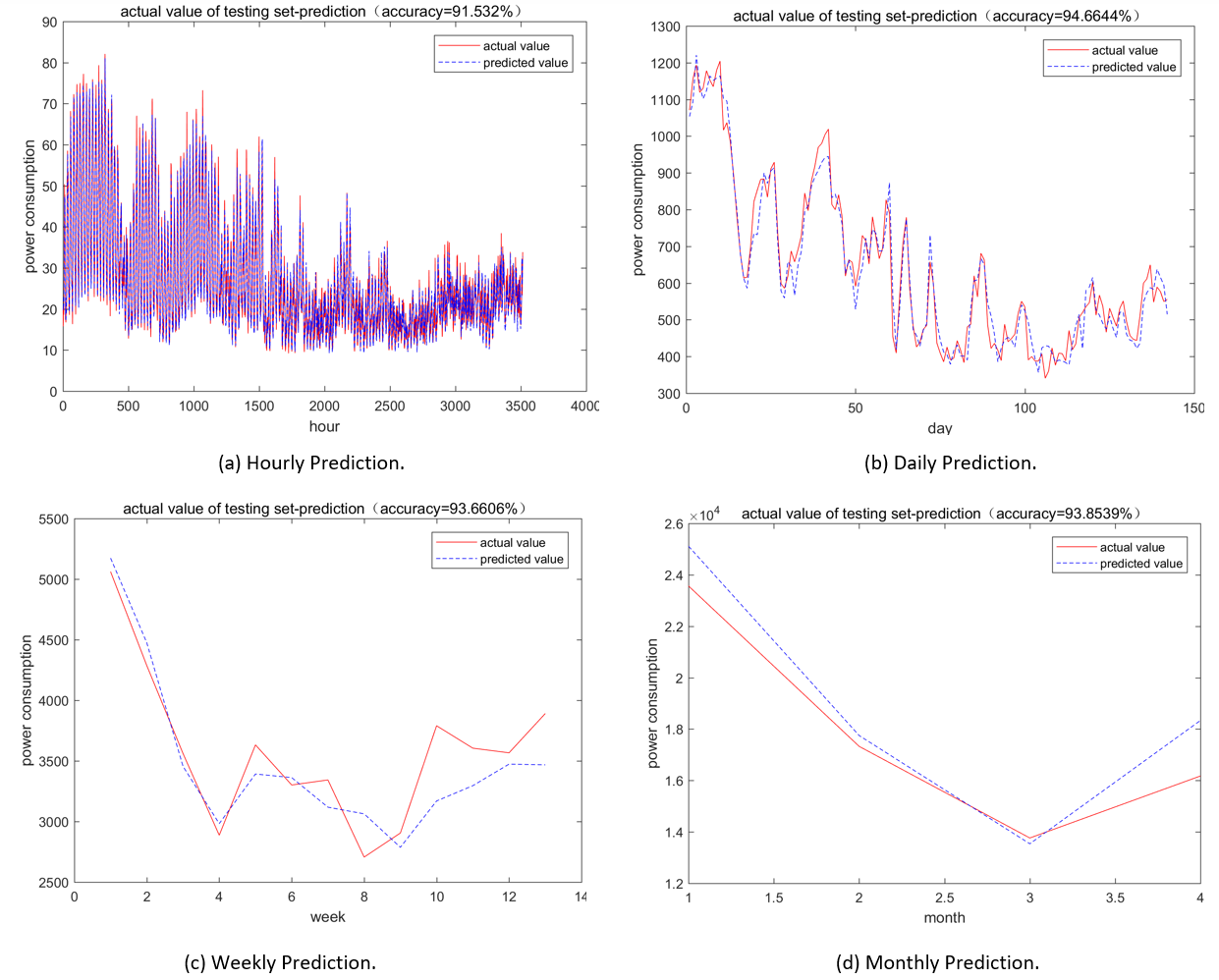}
\end{tabular}
\end{center}
\caption[example] {The best prediction performance with the weather condition factor considered.}
\label{fig:best} 
\end{figure}

\subsection{Discussions}
\subsubsection{Influences of historical record}

As the experimental results presented above, it is clear that taking the weather and weekend/holiday factors into account yields better prediction performance than omitting these factors. However, it is not clear how much historical information should be leveraged to achieve the optimal prediction result. Because of the randomness of computer training and the drawback of the BPNN, the initial weight and bias are generated randomly. The computer training results can easily run into a local optimization solution instead of the global optimization solution. For each experiment, the program runs ten times and the average value is adopted as the ﬁnal result. 

Figure \ref{fig:best} shows the best prediction performance in four time scales with the weather condition factor considered. Among the four prediction time resolutions, the hourly prediction and the daily prediction have achieved better fit than the weekly and the monthly predictions. However, the average accuracy of the monthly and the weekly predictions were slightly higher than the hourly prediction. The average MSE of the hourly and the daily predictions are lower than the weekly and the monthly predictions. However, the prediction performance of the monthly and the weekly predictions are unstable. For example, the accuracy of prediction varies from 69\% to 93\%. The performance of the daily and the hourly predictions is much more stable with the variation below 2\%.

It is not a surprise that this research confirms the data size is the dominant factor that brings impact to the variation in accuracy. The hourly prediction and the daily prediction have much larger training and testing data sets. Under the repeated experiments, the chance factor and local optimum are removed. Theoretically, the higher quality of the training data, the better the prediction performance. Sometimes, there are several data points that deviate from the norm range. But their influences can be mitigated when a sufficiently large data set is available. Meanwhile, if the data set is limited, the abnormal data points will make the prediction result fluctuate wildly, even make the misprediction where the predicted result has the opposite changing trend with actual value. 

\subsubsection{Impacts of each individual factor}
\begin{figure} [htb]
\begin{center}
\begin{tabular}{c}
\includegraphics[height=10cm]{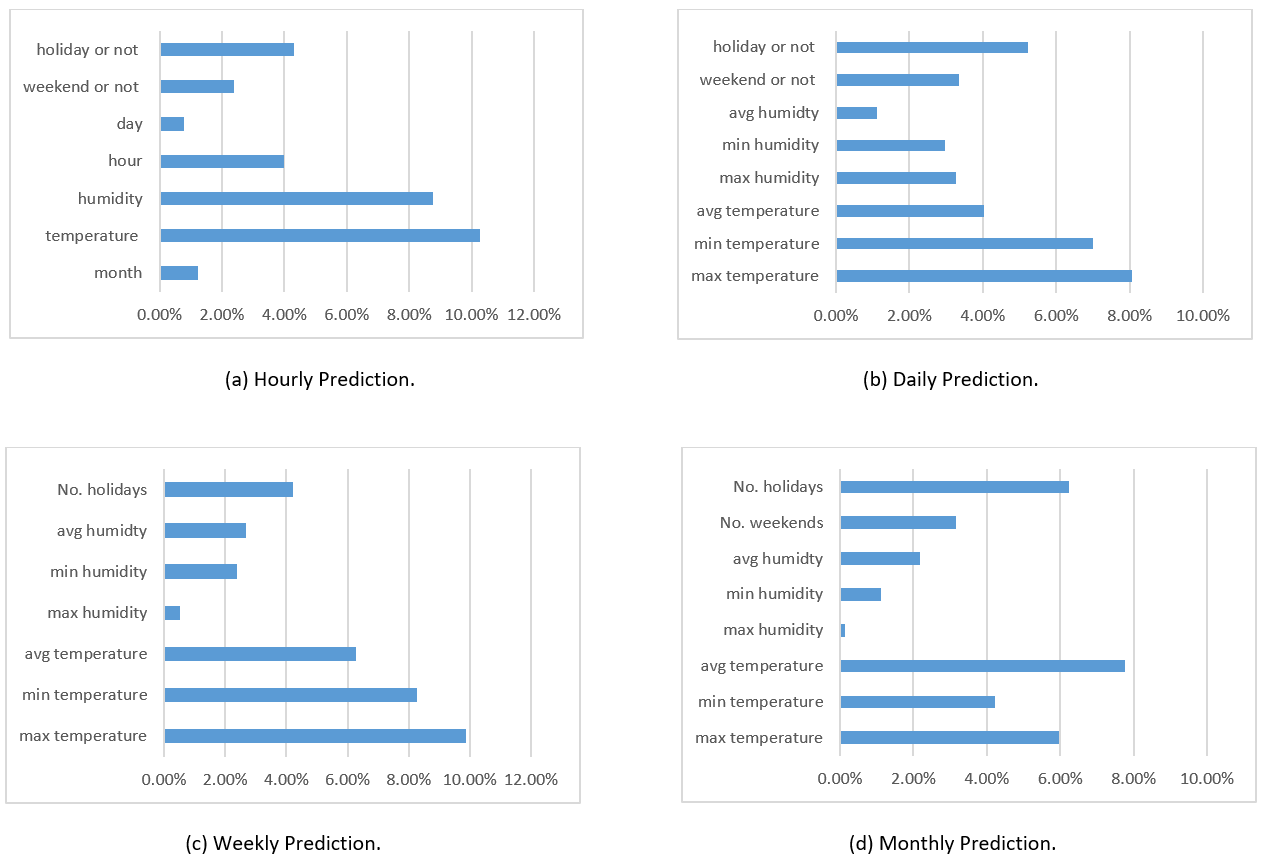}
\end{tabular}
\end{center}
\caption[example] {The influences of each individual factor on the prediction accuracy.}
\label{fig:factors} 
\end{figure}

In this work, the impacts of each individual factor are studied, in all four prediction time scales. Figure \ref{fig:factors} presents the impacts on prediction accuracy when one of the factors is missing. 

\begin{itemize}
    \item \textit{Hourly Prediction:} Fig. \ref{fig:factors}(a) indicates that the temperature and humidity have the largest influences on the prediction performance. However, omitting a given day in a week or a month in a year does not introduce a significant effect in prediction performance.
    \item \textit{Daily Prediction:} as illustrated in Fig. \ref{fig:factors}(b), the highest and the lowest temperatures have the most obvious effect on prediction performance. Meanwhile, the average humidity has the least effect on prediction performance.
    \item \textit{Weekly Prediction:} Figure \ref{fig:factors}(c) shows that the maximum and the minimum temperatures have the most significant influence on the prediction performance. The maximum humidity has the least effect on prediction performance. It is because the maximum humidity maintains above 90\% for many weeks at a stable level, it did not bring much differences in prediction accuracy.
    \item \textit{Monthly Prediction:} as shown by Fig. \ref{fig:factors}(d), it looks all factors are important except the maximum humidity. Because nearly all the months had a maximum humidity of 99\%.
\end{itemize}

\subsubsection{BPNN hidden layer design}

To find the most appropriate number of nodes in the hidden layer, this paper validates empirical formulas, Eq. \ref{eq:exp1} to Eq. \ref{eq:exp4} discussed in Section \ref{sect:proposal}, because there is not a universal formula for prediction models. Therefore, the method of trial and error is utilized. The test case is the daily prediction with the weather and weekend/holiday factors considered, using seven days of historical electricity consumption history. Because this model has relatively stable performance.  For the four different empirical formulas, with 15 input nodes and one output node, the number of nodes in the hidden layer should be as follows:

\begin{itemize}
    \item $\sum_{i=0}^{n} C_{m}^{i} > k$: because the number of samples is larger than 700, the number of nodes in the hidden layer should be larger than 10.
    \item $n_1 = \sqrt{n+m}+a$: the number of nodes in the hidden layer should be around 5 to 14.
    \item $n_1 = \log_2n$: the number of nodes in the hidden layer should be around 4.
    \item $n_1 = 2n+1$: the number of nodes in the hidden layer should be around 31.
\end{itemize}

\begin{figure} [htb]
\begin{center}
\begin{tabular}{c}
\includegraphics[height=5cm]{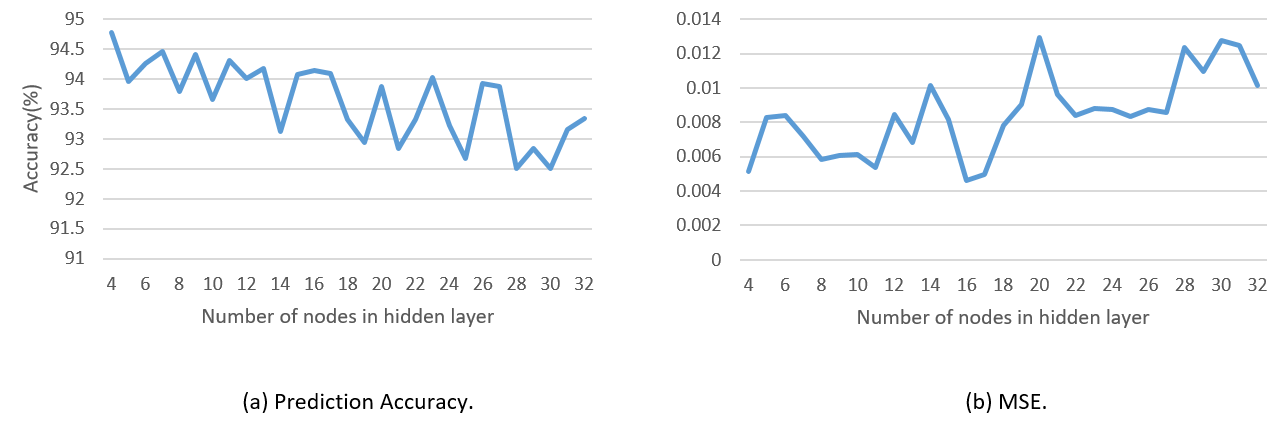}
\end{tabular}
\end{center}
\caption[example] {The influences of different number of nodes in hidden layer.}
\label{fig:hidden} 
\end{figure}

The experimental results are shown in Fig. \ref{fig:hidden}. When the number of nodes in the hidden layer increases, the prediction accuracy appears a constant flux wave, but follows a downward trend. In terms of MSE, the best prediction performance occurs when the number of nodes in the hidden layer is 15 or 16. According to the experimental results, none of the four empirical formulas is accurate. They are concluded based on some earlier experiences. But for different BPNN models, the number of nodes in the hidden layer has to be tried with different values based on an empirical formula, there is not a clear rule.

\section{Conclusions}
\label{sect:conclusions}

In this paper, the BPNN is applied to predict electricity consumption. Leveraging the data set from the Pecan Street Project, the inﬂuences of several factors are experimentally investigated. According to the experimental results, including weather and weekend/holiday factors will increase the prediction accuracy and reduce the MSE. For different prediction timescales, the effects of including weather and weekend/holiday factors are different. For example, the maximum humidity has less effects on weekly and monthly predictions than it has for hourly and daily predictions. Hourly and daily predictions have relatively better performance than weekly and monthly predictions. One of the most important reasons could be the size of the data set. Theoretically, the better quality the training data possesses, the higher the accuracy. Regarding the BPNN, there are many empirical formulas or commonly used parameters, such as learning rate, epoch times, the proportion of different data sets (e.g., training, testing, and validation data sets), the number of nodes in the hidden layer and so on. However, when conducting real experiments, we have to repeat experiments to ﬁnd the best value because there are not well-deﬁned guidelines for choosing the parameters.



\bibliography{report}   
\bibliographystyle{spiejour}   

\end{document}